\documentstyle[aps,twocolumn]{revtex}
\input{psfig}

\begin{document}
\draft
\def\be{\begin{equation}}
\def\ee{\end{equation}}
\def\beq{\begin{eqnarray}}
\def\eeq{\end{eqnarray}}
\wideabs{
\title{Wave Function of the Largest Skyrmion on a Sphere}
\author{Daijiro Yoshioka}
\address{Department of Basic Science, The University of Tokyo\\
3-8-1 Komaba, Meguro-ku, Tokyo 153-8902, Japan
}
\date{June 26, 1998}
\maketitle
\begin{abstract}
It has been clarified that charged excitation known as a skyrmion exists around the ferromagnetic ground state at the Landau level filling factor $\nu=1/q$, where $q$ is an odd integer.
An infinite sized skyrmion is realized in the absence of the spin-Zeeman splitting or for double-layered systems.
Analytical form of the wave function is identified at $\nu=1$ and $\nu=1/3$ through exact diagonalization of the Hamiltonian for finite sized spherical systems.
It is clarified that the skyrmion wave functions at $\nu=1$ and $\nu=1/3$ are qualitatively different: they are not related by the composite fermion transformation.
Long-range behavior of the skyrmion wave function around $\nu=1$ is shown to be consistent with the semiclassical picture of the skyrmion.
\end{abstract}
\pacs{Keywords: FQHE, Skyrmion, composite fermion, spin-singlet ground state
}
}

\narrowtext

Investigations of the quantum Hall effect are mainly carried out on GaAs-AlGaAs heterostructures.
Recently it has been clarified that due to the small g-factor of electrons in this material, electronic spin freedom sometimes plays an important role even in strong magnetic fields.
One such phenomenon is the so-called skyrmion excitation around $\nu=1$, where $\nu$ is the Landau level filling factor.\cite{barret,schmeller,aifer}
Namely, although the electron spins are aligned ferromagnetically at $\nu=1$, removal or addition of an electron causes spin flip of other electrons.\cite{sondhi,fertig}
In a typical experiment three spin flips per hole or electron have been observed.

The number of spin flips depends on the ratio of the spin Zeeman energy and the strength of the exchange interaction between electrons.
In the limit of vanishing Zeeman splitting, it is expected that half of the electrons flip their spins and a spin singlet ground state is realized.
Actually, Rezayi showed that such a marked change of the ground state occurs for electron systems on a sphere.\cite{rezayi,rezayi1}
He found that if there is a spin degeneracy, the ferromagnetic ground state at $\nu=1$ changes into the spin singlet ground state by the addition or removal of an electron.

\begin{figure}
\psfig{figure=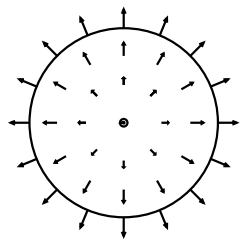,height=4cm}
\caption{Classical skyrmion on a sphere.
The electrons spins are aligned locally and point radially outwards.
\label{fig:1}}
\end{figure}

The origin of this phenomenon is the ferromagnetic exchange interaction between electrons.\cite{moon}
The ground state at $\nu=1$ is ferromagnetic even if the Zeeman splitting is zero.
If one more electron is introduced, its spin should be opposite to the others to avoid the kinetic energy.
Then to lower the exchange interaction the introduced electron causes the spin flip of surrounding electrons so that spins are locally aligned.
Therefore the classical picture of the total spin zero state resembles that shown in Fig.~\ref{fig:1}.
Due to the spin-charge relation\cite{lee,sondhi} this spin texture is accompanied by charge such that the extra electron can be accommodated in the lowest Landau subband.
Because of the electron-hole symmetry at $\nu=1$ the introduction of a hole also creates spin texture similar to that in Fig.~\ref{fig:1} except that all the spins are reversed.

However, quantum mechanically the spin texture in Fig.~\ref{fig:1} does not represent the spin singlet state.
This is obvious since the state in Fig.~\ref{fig:1} is degenerate with respect to global spin rotation, but the quantum mechanical spin singlet state is nondegenerate.
The aim of the present study is to clarify the quantum mechanical wave function of the spin singlet ground state for the spherical system without Zeeman splitting.
Since the situation is quite similar at $\nu=1/3$, we shall also investigate the spin singlet ground states around $\nu=1/3$.
We, hereafter, call the spin singlet ground state realized on a sphere by reducing one flux quantum from the $\nu=1/q$ state as qe-skyrmion (quasielectron-skyrmion), and that by introducing one more flux quantum into the $\nu=1/q$ state as qh-skyrmion state.

The ferromagnetic ground state wave function at $\nu=1/q$ is quite well approximated by the Laughlin wave function.\cite{laughlin}
Halperin generalized it by considering spin degrees of freedom, and wrote down a series of wave functions now known as the $\Psi_{m,m,n}$ wave function:\cite{halperin}
\beq
\Psi_{m,m,n} &=& \prod_{i>j}(u_iv_j-v_iu_j)^m(\xi_i\eta_j-\eta_i\xi_j)^m
\nonumber\\
&\times&\prod_{i,j} (u_i\eta_j-v_i\xi_j)^n,
\label{eq:1}
\eeq
where $(u_i,v_i)=(\cos(\theta_i/2){\rm e}^{{\rm i}\phi_i/2},\sin(\theta_i/2) {\rm e}^{-{\rm i}\phi_i/2})$ and $(\xi_i,\eta_i)$ are complex spinor coordinates of spin-up and spin-down electrons on the sphere, respectively.\cite{haldane}

When the number of spin-up and spin-down electrons are the same, Halperin's wave function results in a state with total $S_z=0$.
However, it has been pointed out\cite{hr} that except for the case of $m=n+1$, Halperin's wave function does not represent valid orbital part of the spin singlet wave function, since it does not satisfy the Fock condition.\cite{hamer}
Actually, for $m=n$ it still represents a maximally spin polarized ferromagnetic state even though total $S_z=0$.

Haldane and Rezayi modified the Halperin wave function to discuss the spin singlet state around $\nu=1/2$.\cite{hr}
Later, Yoshioka et al. found that around $\nu=1/q$ the following wave function with $r=1$ satisfies the Fock condition and represents the spin singlet wave function:\cite{yoshioka}
\be
\Psi_q^{(r)} = \Psi_{q,q,q}{\rm per}|M^{(r)}|,
\label{eq:2}
\ee
where for an $N_e$ electron system $M^{(r)}$ is a $N_e/2 \times N_e/2$ matrix with matrix elements $M_{ij}^{(r)}=(u_i\eta_j-v_i\xi_j)^r$, and per$|M|$ implies the permanent of the matrix $M$.
By simply counting the maximum power of the electron coordinate $u_i$ or $\xi_i$, we find that the filling factor of this wave function is such that $r$ extra flux quanta are introduced into the system at $\nu=1/q$.
Since, it is easily verified that this wave function satisfies the Fock condition for the spin-singlet wave function for any {\it odd} integer $r$, the wave function for $r=-1$ was proposed by Rezayi as a trial wave function for spin-singlet state at such a filling factor.\cite{rezayi1}
In this paper we examine if this series of wave functions give good approximation to the spin singlet wave function on a sphere or not.

Since the wave function, eq.~(\ref{eq:2}), is not a unique ground state of any model Hamiltonian, which is described by the Haldane pseudopotential,\cite{haldane}
the wave function in the second quantized form was constructed by expanding eq.~(\ref{eq:2}) using Mathematica.
The overlap of the obtained wave function with the Coulomb ground state wave function obtained by the exact diagonalization method has been calculated.
The results are shown in Tables I and II for the skyrmions around $\nu=1$ and $\nu=1/3$, respectively.
The results show that around $\nu=1$ the qh-skyrmion is almost perfectly approximated by the model wave function $\Psi_1^{(1)}$, but the qe-skyrmion is not well approximated by the model wave function $\Psi_1^{(-1)}$.
On the other hand, around $\nu=1/3$ the opposite is true: the qe-skyrmion is quite well approximated by $\Psi_3^{(-1)}$, but the qh-skyrmion is not by $\Psi_3^{(1)}$.

At $\nu=1$ the electron-hole symmetry is present.
Therefore, the qe-skyrmion state is essentially identical to the qh-skyrmion state.
This means that the qe-skyrmion state is well approximated by the electron-hole symmetric state of $\Psi_1^{(1)}$, which we denote as $\tilde\Psi_1^{(1)}$.
Even though the filling factors at which they are realized are the same, the model wave function $\Psi_1^{(-1)}$ is not the same as $\tilde\Psi_1^{(1)}$.
The two states compete for lower energy and that with $r=1$ wins.

On the other hand, there is no electron-hole symmetry around $\nu=1/3$.
Therefore, there is no relation between the qe-skyrmion state and the qh-skyrmion state.
For the qe-skyrmion state, one candidate is the model wave function $\Psi_3^{(-1)}$.
However, just as $\Psi_1^{(-1)}$ and $\Psi_3^{(-1)}$ are related by the composite fermion transformation:
\be
\Psi_3^{(-1)} = D_2\Psi_1^{(-1)},
\label{eq:21}
\ee
where
\be
D_2 = \prod_{i>j}(u_iv_j-v_iu_j)^2(\xi_i\eta_j - \eta_i\xi_j)^2
\prod_{i,j}(u_i\eta_j-v_i\xi_j)^2,
\label{eq:22}
\ee
so that $D_2 \tilde\Psi_1^{(1)} \equiv \tilde \Psi_3^{(1)}$ can be another candidate.
For the present filling factor, in contrast to the case of $\nu=1$, $\Psi_3^{(-1)}$ approximates the true ground state well.

Similarly, there are two candidates for the $\nu=1/3$ qh-skyrmion state.
One is $\Psi_3^{(1)}$ and the other is $D_2\tilde \Psi_1^{(-1)} \equiv \tilde \Psi_3^{(-1)}$, where $\tilde \Psi_1^{(-1)}$ is the electron-hole symmetric state of $\Psi_1^{(-1)}$.
Table II shows that $\Psi_3^{(1)}$ does not provide a good approximation to the ground state.
It may be possible that $\tilde \Psi_3^{(-1)}$ is realized here, but we have no evidence at present.
It was found that the overlap improves to 0.99967 for the four-electron system by slightly reducing the pseudopotential for relative angular momentum 3, $V_3$, from the Coulomb value of 0.28252 to 0.27.
Further reduction of $V_3$, however,
produces an almost zero overlap.
This is because $V_3$ approaches $V_4=0.25427$ too closely.
Thus model pseudopotentials, where $V_0$, $V_1$, and $V_2$ are quite large, $V_3$ is finite and all the other higher $V_m$'s are zero, also give an almost exact model wave function, $\Psi_3^{(1)}$.\cite{note2}

It should be noted that because of the existence of two almost degenerate ground states represented by $\Psi_q^{(\pm 1)}$ and $\tilde \Psi_q^{(\mp 1)}$, the single-skyrmion ground state depends on the relative weights of the pseudopotentials.
The ground state around $\nu=1$ is not related to that around $\nu=1/3$ by the simple composite fermion transformation.

Having identified the quantum mechanical wave function of the skyrmion state, let us consider the relation to the semiclassical picture of the skyrmion.
As we have seen before, the classical picture shown in Fig.~\ref{fig:1} is not exact from the quantum mechanical point of view.
The skyrmion in Fig.~\ref{fig:1} is represented in quantum mechanics by a Hartree-Fock wave function given in refs.~5 and 8.
The single electronic states are chosen as eigenstates of the total angular momentum ${\bf J}={\bf L}+{\bf S}$, where ${\bf L}$ is the orbital angular momentum of the electron on the sphere, and ${\bf S}$ is its spin.
When $2l$ flux quanta pass through the surface of the sphere, $L_z$ takes values between $-l$ and $l$.
When these $2l+1$ states are occupied by parallel spin electrons, $\nu=1$ ferromagnetic state is realized.
The Hartree-Fock qh-skyrmion state is the state in which all of the $J=l-1/2$ single electron states are occupied.
This state is the eigenstate of the total angular momentum, the eigenvalue being zero.
However, it is not the eigenstate of the total orbital angular momentum or the total spin angular momentum.
Therefore, the overlap of this state with the true ground state skyrmion wave function is quite small: for a four-electron system it is only 0.4289.
Even if the Hartree-Fock wave function is projected onto the $S=0$ and $L=0$ state, the overlap improves only up to 0.6538.
Therefore, the semiclassical Hartree-Fock wave function is not a good approximation of the true ground state.

In spite of the difference in the wave function, the long-range spin-spin correlation seems to be quite accurately described by the semiclassical picture at $\nu=1$.
In order to study this property, we calculated the two-particle correlation function $g_{\sigma, \sigma'}(\theta)$, which is defined as
\be
g_{\sigma,\sigma'}(\theta) = \frac{S}{N_\sigma N_{\sigma'}}
\frac{1}{2\pi\sin\theta}
\langle \sum_{i \ne j} \delta(\theta - \theta_{i,\sigma,j,\sigma'})\rangle,
\label{eq:3}
\ee
where $S$ is the surface area of the sphere, $N_{\sigma}$ is the number of electrons with spin $\sigma$, $\theta_{i,\sigma,j,\sigma'}$ is the angle between $i$-th electron with spin $\sigma$ and $j$-th electron with spin $\sigma'$: $\cos\theta_{i,\sigma,j,\sigma'} = \cos\theta_{i,\sigma}\cos\theta_{j,\sigma'} +\sin\theta_{i,\sigma}\sin\theta_{j,\sigma'} \cos(\phi_{i,\sigma}-\phi_{j,\sigma'})$, and the angular bracket indicates the expectation value with respect to the ground state or $\Psi_q^{(r)}$.
The results for the Coulomb ground state with twelve electrons are shown in Fig.~2(a) and Fig.~2(b) for the qh-skyrmion state and the qe-skyrmion state, respectively.
In these figures the thick solid (dashed) line shows $g_{\uparrow,\uparrow}(\theta)$ ($g_{\uparrow,\downarrow}(\theta)$).
The thin dotted lines show the correlation function for the semiclassical skyrmion shown in Fig.~1, which can be analytically calculated to be
\be
g_{\uparrow,\uparrow}(\theta) = \frac{2}{3}[1+ \cos^2(\frac{\theta}{2})],
\label{eq:4}
\ee
\be
g_{\uparrow,\downarrow}(\theta) = \frac{2}{3}[1+ \sin^2(\frac{\theta}{2})].
\label{eq:5}
\ee

Comparison of the quantum mechanical and semiclassical results clearly shows that the long-range correlation is well approximated by the semiclassical picture.
For the qh-skyrmion state, the difference is larger.
However, the size dependence is larger here, and the probability to find the same (opposite) spin electron at $\theta=\pi$ tends to 2/3 (4/3) for the infinite size system, which is consistent with the semiclassical results.\cite{note1}

On the other hand, short-range behavior is quite different.
Of course most of the difference in $g_{\uparrow,\uparrow}$ around the origin comes from the exchange hole, which cannot be taken into account in the semiclassical treatment.
We should note, however, that short-range repulsion also exists between the opposite spin electrons.
The short-range repulsion is stronger for the skyrmions around $\nu=1/3$.
Since the difference should be confined in the range of the exchange hole, we can expect that for an infinitely large system, the spin-spin correlation
can be explained by the semiclassical picture in most areas.
\begin{figure}
\psfig{figure=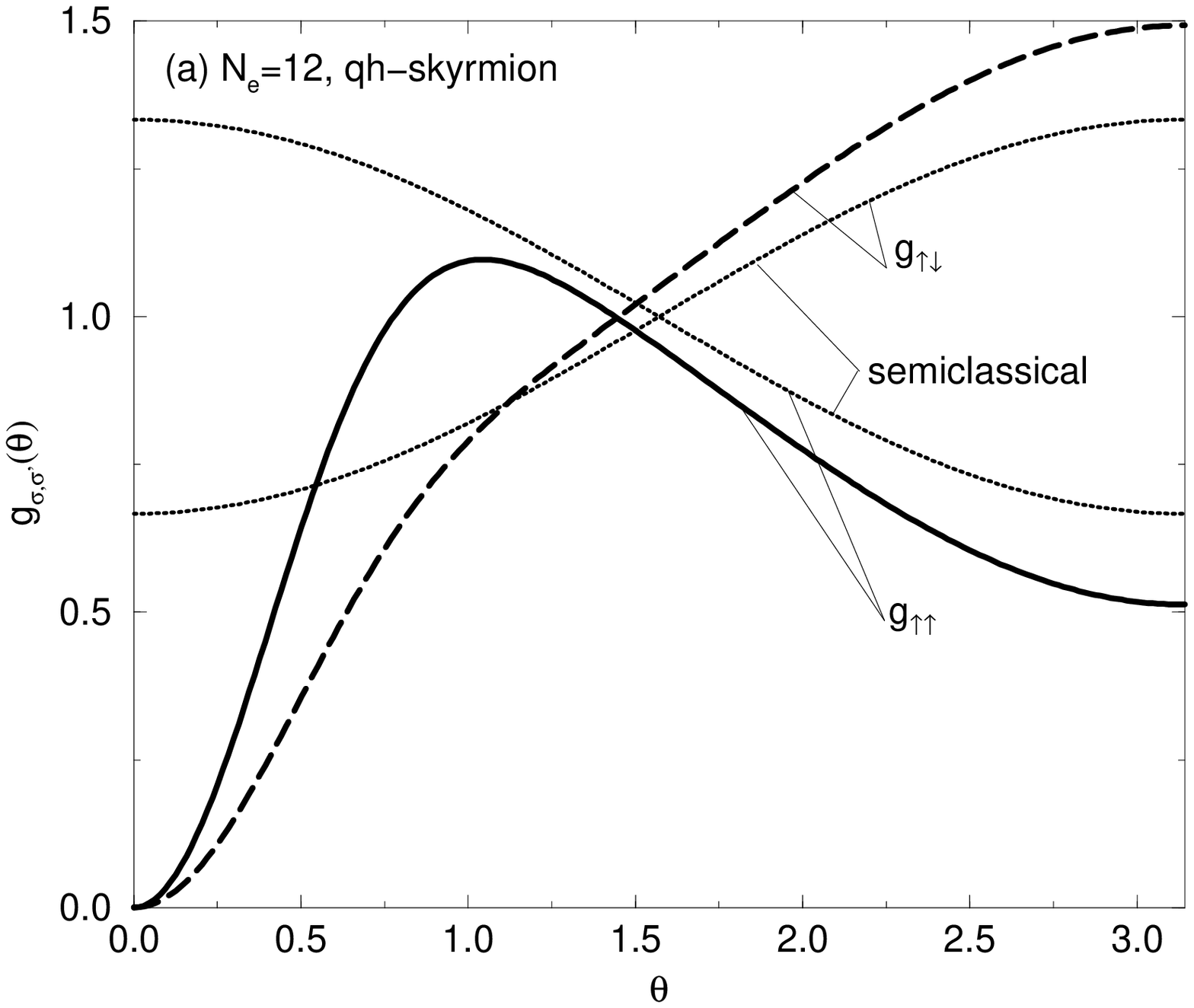,height=7cm}
\end{figure}
\begin{figure}
\psfig{figure=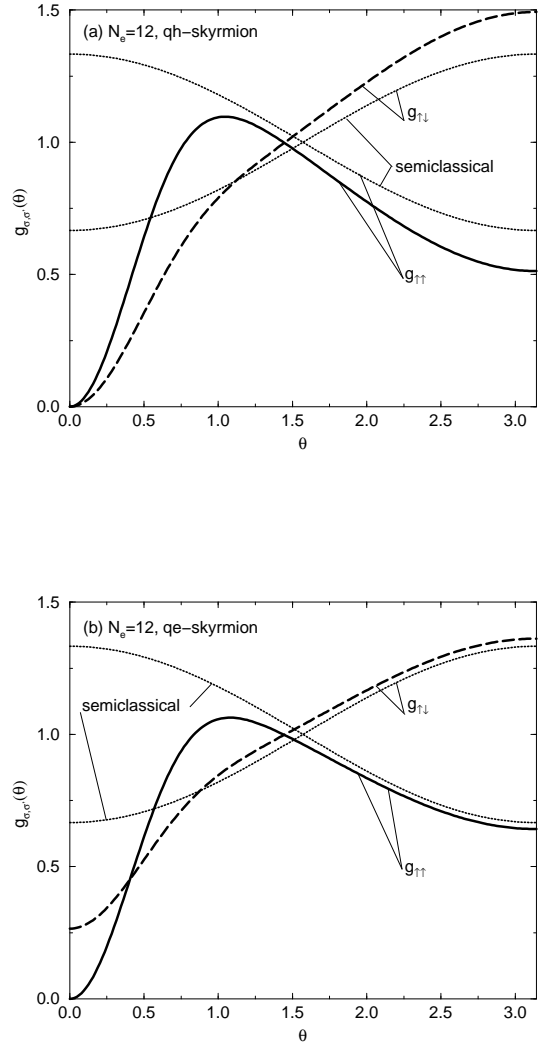,height=7cm}
\caption{Two-particle spin correlation function $g_{\sigma,\sigma'}$ of (a) the qh-skyrmion and (b) the qe-skyrmion state around $\nu=1$.
The thick solid and dashed lines show $g_{\uparrow,\uparrow}$, and  $g_{\uparrow,\downarrow}$ for a twelve-electron system, respectively. The thin dotted lines are those for the semiclassical skyrmion.
\label{fig:2}}
\end{figure}

It is important to notice that the good agreement in the long-range behavior is obtained because the Coulomb ground state is described by $\Psi_1^{(1)}$ or  $\tilde\Psi_1^{(1)}$, and not by $\Psi_1^{(-1)}$.
In Fig.~3, $g({\bf r})$'s for $\tilde\Psi_1^{(1)}$ and $\Psi_1^{(-1)}$ are compared for the 8-electron system.
The overlap of these wave functions is 0.94072, but $g({\bf r})$'s show noticeable differences, especially in the long-range behavior.
The correlation function for the true Coulomb ground state coincides with that of
$\tilde\Psi_1^{(1)}$ within the thickness of the curves, and is consistent with the semiclassical results similar to those of the 12-electron system.
Thus we have established that except for the short-range correlation, the semiclassical skyrmion is consistent with the true skyrmion ground state around $\nu=1$, and quantum mechanically it is described by $\Psi_1^{(1)}$.
\begin{figure}
\psfig{figure=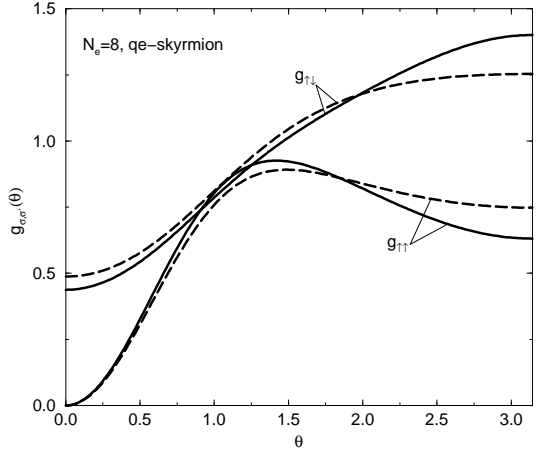,height=7cm}
\caption{Comparison of $g_{\sigma,\sigma'}$ for $\tilde\Psi_1^{(1)}$ and $\Psi_1^{(-1)}$.
The solid lines show $g_{\sigma,\sigma'}$ for $\tilde\Psi_1^{(1)}$, and the dashed lines show $g_{\sigma,\sigma'}$ for $\Psi_1^{(-1)}$. Here the number of electrons is 8.
\label{fig:3}}
\end{figure}

%
%
In this paper the wave functions of the largest skyrmions on a sphere are identified.
The extent to which the semiclassical picture gives a good description of the true Coulomb ground state has been clarified.
It is expected that the difference in the short-range spin-spin correlation between the semiclassical and quantum descriptions becomes more important for finite sized skyrmions realized experimentally.

{\it Note added} ---
After this paper was submitted, it is informed by J. Jain that their
group has also investigated the skyrmion states around $\nu=1/3$.\cite{jain1,jain2,jain3}
From comparison of the energy they arrived at a conclusion that the skyrmions around $\nu=1/3$ are related to those around $\nu=1$ by the composite fermion transformation.
We remark that our conclusion in the present study which contradicts theirs is obtained by comparison of the wavefunctions which is more stringent than the comparison of the energies.

\begin{table}
\caption{Overlap of the qh- and qe-skyrmion wave functions realized around $\nu=1$ with $\Psi_1^{(1)}$ and $\Psi_1^{(-1)}$, respectively.}
\label{table:1}
\begin{tabular}{@{\hspace{\tabcolsep}\extracolsep{\fill}}ccc} \hline
Number of electrons & $\langle\Psi_1^{(1)}|{\rm qh}\rangle$ & $\langle\Psi_1^{(-1)}|{\rm qe}\rangle$ \\ \hline
4   & 0.99949 & 0.94281 \\
6   & 0.99910 & 0.97646 \\
8   & 0.99882 & 0.95337 \\ \hline
\end{tabular}
\end{table}
\begin{table}
\caption{Overlap of the qh- and qe-skyrmion wave functions realized around $\nu=1/3$ with $\Psi_3^{(1)}$ and $\Psi_3^{(-1)}$, respectively.
Here $\Psi_3^{(1)}$ for 6 electrons is constructed approximately by the diagonalization of a model Hamiltonian.}
\label{table:2}
\begin{tabular}{@{\hspace{\tabcolsep}\extracolsep{\fill}}ccc} \hline
Number of electrons & $\langle\Psi_3^{(1)}|{\rm qh}\rangle$ & $\langle\Psi_3^{(-1)}|{\rm qe}\rangle$ \\ \hline
4   & 0.96483 & 0.99633 \\
6   & (0.972) & 0.98919 \\ \hline
\end{tabular}
\end{table}


\begin{references}
\bibitem{note2} The wave function $\Psi_3^{(1)}$ for 6 electron system is constructed by this set of model potentials, and used to calculate the overlap in Table II.
\bibitem{note1} Notice that the qh-skyrmion state shown in Fig.~2(a) is the electron-hole symmetric state of the qe-skyrmion state of 14 electrons system.
\end{references}
\end{document}